\documentclass[3p]{elsarticle}

\usepackage{setspace}
\usepackage{amsmath, amsthm, amssymb, mathrsfs}
\usepackage{booktabs}
\usepackage{graphicx}
\usepackage[hidelinks]{hyperref}
\usepackage{pdflscape}
\usepackage{pdfpages}

\makeatletter
\def\ps@pprintTitle{%
   \let\@oddhead\@empty
   \let\@evenhead\@empty
   \let\@oddfoot\@empty
   \let\@evenfoot\@oddfoot
}
\makeatother

\biboptions{authoryear}
\newpageafter{abstract}

\def\sym#1{\ifmmode^{#1}\else\(^{#1}\)\fi}

\begin{document}

\begin{frontmatter}

\title{The geography of novel and atypical research}

\author[1]{Qing Ke\corref{a}}
\author[2]{Tianxing Pan}
\author[3]{Jin Mao}

\address[1]{Department of Data Science, City University of Hong Kong, Hong Kong, China}
\address[2]{School of Information Management, Nanjing University, Nanjing, Jiangsu Province, China}
\address[3]{Center for Studies of Information Resources, Wuhan University, Wuhan, Hubei Province, China}
\cortext[a]{Corresponding author: q.ke@cityu.edu.hk}

\begin{abstract}
The production of knowledge has become increasingly a global endeavor. Yet, location related factors, such as local working environment and national policy designs, may continue to affect what kind of science is being pursued. Here we examine the geography of the production of creative science by country, through the lens of novelty and atypicality proposed in \citet{Uzzi-atypical-2013}. We quantify a country's representativeness in novel and atypical science, finding persistent differences in propensity to generate creative works, even among developed countries that are large producers in science. We further cluster countries based on how their tendency to publish novel science changes over time, identifying one group of emerging countries. Our analyses point out the recent emergence of China not only as a large producer in science but also as a leader that disproportionately produces more novel and atypical research. Discipline specific analysis indicates that China's over-production of atypical science is limited to a few disciplines, especially its most prolific ones like materials science and chemistry.
\end{abstract}


\end{frontmatter}

\section{Introduction}

Originality and novelty are widely appreciated in scientific research. It is every researcher's efforts to make new contributions to the knowledge base in a field of study. Novel research, however, is considered as high-risk and high-return; it features a greater potential for scientific breakthroughs but bears a higher risk of failure. Manifestations of risks include experiencing resistance of acceptance from established scientific paradigms and delayed appreciation and recognition of its importance from the scientific community. Empirical studies have indeed found supporting evidence for the risk profile of novel science, showing that there is a bias against novel research. Novel scientific papers are more likely to exhibit delayed recognition and get published in low Impact Factor journals \citep{WANG20171416}; novel research proposals venturing into uncharted territories are discouraged from increasingly risk-averse funding agencies, which favor relatively safer projects \citep{Boudreau2016looking, wang2018funding}; and novel ideas proposed by demographically underrepresented doctoral students are less likely to be taken up \citep{Hofstra9284}. Possibly due to these factors, scientific novelty has been a minority in published works \citep{Uzzi-atypical-2013, Ke2020104071, Rzhetsky14569}, and researchers have adopted conservative research strategies \citep{Foster2015tradition}.

In innovation studies, a number of ways have introduced to identify novel research. They are predominately from the knowledge recombination perspective, resting on the widely accepted sociological theory that the production of new knowledge is resulted from a search and recombination process over existing knowledge components \citep{schumpeter1939business}. Consequently, novel knowledge is considered to recombine prior knowledge that has not been recombined before. Despite different operationalization methods, extant studies have reached the same conclusion that novel combinations of existing knowledge components result in scientific breakthroughs. A unique insight from \citet{Uzzi-atypical-2013} is that novel combination of preceding works, when joined with conventional one, results in atypical advances that are more likely to be scientific breakthroughs.

Building on this premise of examining creativity in science through the lens of combining novelty and conventionality, in this work we study the geography dimension of the production of novel and atypical research. Specifically, our research questions are:
\begin{enumerate}
\item How do different countries perform in producing novel and atypical science, and how does the performance change over time?
\item What are the roles of discipline and international collaboration in explaining performance differences between countries?
\end{enumerate}
Here our hypothesis is that, although knowledge production has become increasingly a global endeavor, location may continue to play an important role in what kind of science is pursued. This is because location may well affect the tendency for scientists to conduct novel science in several ways. First, local working environment, such as interaction with and access to colleagues and collaborators with diverse background, support from mentors, and risk-tolerant environment, is crucial for fruitful production of creative work. Second, many important science policy designs are made at the national level, which may also have considerable influence on the fertility of novel science. Unraveling spatial distributions of novel research would pave the way to identifying potential barriers to pursue novel research and to further designing policy interventions to eliminate such barriers.

Using the Microsoft Academic Graph (MAG) dataset containing almost 38 million papers \citep{Wang2019MAG}, we first present an excellent  reproducibility about the relationship between atypicality and scientific impact, which serves as a validation of our implementation of the calculation of novelty and conventionality. We then examine countries where novel and atypical research are produced, finding that differences in the tendencies of novel and atypical science production continue to persist among countries. Clustering analysis of countries based on their propensity to produce novel science over time reveals groups of countries that are emerging, including China, Korea, etc. Our analysis points out the rise of China not only as one of the largest producers in science but also as a top country that disproportionately yields more novel and atypical papers. Further discipline specific analyses reveal heterogeneity in country performances. For example, the US over-produced atypical papers in all the disciplines, whereas the over-production of atypical science from China was limited to materials science, chemistry, and computer science. Meanwhile, countries tend to have similar performances for the two groups of papers based on whether they involve international collaborations. Finally, to facilitate followup inquiry related to novelty, we have made publicly available the data.

\section{Related literature} \label{sec:lit}

\subsection{Knowledge recombination and combinatorial novelty}

There is a long tradition to conceptualize novelty from a knowledge recombination perspective. Back in the 1930s, Schumpeter emphasized the importance of recombination on business cycle and stated the recombinant nature of innovation, arguing that innovation is the recombination of resources such as technology, capital, and people in new ways \citep{schumpeter1939business}. Viewing knowledge creation as a search process that recombines existing knowledge elements, the combinatorial perspective of novelty theorizes novelty as unusual recombinations of antecedent knowledge. This theory has been embraced by scholars in a number of disciplines over the last few decades \citep{nelson1985evolutionary, henderson1990architectural, kogut1992knowledge, fleming2001recombinant}. In the technological innovation literature, for instance, \citet{fleming2001recombinant} used technology subclasses to represent technological knowledge components and considered more novel combinations of subclasses as those less familiarized by an inventor, defined through their appearances in the inventor's previous patents. He found that novel combinations are linked to lower average but higher variance of forward citations, highlighting the impact uncertainty brought by novelty. Along similar lines, \citet{verhoeven2016measuring} provided a more refined classification of novelty of patents, including novelty in technological and scientific knowledge sources, based on whether a patent is the first in its patent class to cite the literature from domains that have never been cited before. They found that the combination of novelty in both aspects is associated with breakthrough inventions.

Turning to the scientific novelty literature, conceptually similar methods have been proposed for novelty quantification of scientific articles. \citet{Uzzi-atypical-2013} viewed journals as bodies of knowledge ingredients and calculated the unusualness of each journal pair as the relative deviation of its observed frequency from random expectations, expressed as a $z$-score. A paper's conventionality and novelty are then respectively determined by 50th- and 10th- percentile of its $z$-score distribution of cited journal pairs. This method was validated by subsequent researchers, though not based on the exact replication and using other data sources like Scopus \citep{boyack2014atypical, Wagner-intl-2019}. Using a modified version of the novelty measure from \citet{Uzzi-atypical-2013}, \citet{lee2015creativity} unpacked the differentiated effect of team characteristics on novelty from the effect on impact. \citet{WANG20171416} developed a novelty measure by operationalizing uncommonness of journal pairs as cosine distance between them, and demonstrated potential bias against novel research when using widely adopted bibliometric indicators, by showing that novel research is less likely to become highly cited in a short time windows and is published in lower Impact Factor journals. Other studies have traced the link between science and technology to explore the impact of scientific novelty on technological impact \citep{veugelers2019scientific, Ke2020104071}.

Although novelty quantification in the literature has in general followed the combinatorial perspective, a number of lines of variations have been introduced. First, knowledge components have been variously operationalized as journals \citep{Uzzi-atypical-2013, WANG20171416}, chemicals \citep{Foster2015tradition, Rzhetsky14569}, predefined keywords \citep{Boudreau2016looking, Ke2020104071}, scientific concepts as detected in text \citep{Hofstra9284}, and topics derived from topic modeling \citep{Kaplan2015double}. Second, two broad categories of variations have emerged when measuring the degree of unusualness. The first one views first-ever combination as a sign of unusualness \citep{Ke2020104071, Boudreau2016looking, Hofstra9284, verhoeven2016measuring}, whereas the other takes frequency of combinations into consideration \citep{fleming2001recombinant, Uzzi-atypical-2013, lee2015creativity, WANG20171416}. Third, while the majority of studies have focused on individual pairs of knowledge components, some scholars measured novelty of 3-tuple knowledge combinations \citep{keijl2016two}. Other studies have assessed novelty from a network perspective, formalizing the knowledge discovery process as uncovering links in the web of knowledge. \citet{Rzhetsky14569} and \citet{Foster2015tradition} built a network of chemicals and proposed a categorization of biomedical research strategies based on how papers explored chemical knowledge networks.

Finally, an important distinction between novelty and impact is worth mentioning. Novelty is an \emph{ex ante} characteristic that is determined at the point when the focal scientific article is produced, capturing whether it makes new combinations of existing ideas, whereas impact is related to \emph{ex post} recognition that is achieved through the social interaction process of the paper in the academic community, emphasizing the subjective judgment of the scientific publication. As such, novelty and impact may not be achieved simultaneously, and there is a problem of difficulty in recognition and delayed identification. Highly novel studies may be difficult to be appreciated due to potential resistance from existing paradigms \citep{veugelers2019scientific}. One aim of extant research is indeed to unravel the relationship between novelty and impact. \citet{trapido2015novelty} concluded that the acceptance of novel innovation is influenced by the characteristics of the audience, such as audience's age, cultural differences, and perception of adverse effects of novelty. Furthermore, novel studies are more likely to become delayed recognition \citep{WANG20171416}; they may not receive significantly higher citations in the short term but are more likely to become highly cited in the long term, reflecting its high-risk and high-return features. At the author level, novel research does not always bring its authors the corresponding acclaim from the scientific community in the short term, but a continued publishing of innovative research can compensate the negative effect of exploration \citep{trapido2015novelty}.

\subsection{Novelty and science policy}

Novelty has also been one of the focal points in recent science policy discussions, particularly on the relationship between novelty, risk-taking, and science funding. Over an extended period of time, one goal of formulating, enacting and adapting science policy have been to stimulate scientists across fields to conduct productive and pioneering research. Yet, many commentaries and empirical studies point out the decline of innovative research, as measured in diverse ways~\citep{park2023papers, Foster2015tradition}. The literature has frequently noted two interrelated contributing factors that may potentially curtail the inclination of researchers to engage in novel research, thereby impeding the progress of knowledge frontiers \citep{stephan2017reviewers}. The first factor is that science funding system is increasingly risk-averse in selecting proposals, remaining instead attuned to the feasibility of proposed research. Viewing novelty as one dimension of risk-taking, \citet{veugelers2022funding} focuses on the European Research Council funding program and offered empirical evidence supporting selection penalty for high-risk research. They found that applicants with publication records of risky research encounter diminished success in securing funding than those without such records, a trend that holds for early career researchers as well as researchers with highly cited papers. The study also evaluated the effect of receiving grants in promoting risk taking, finding a higher likelihood for early career recipients to engage in risky research when compared with unsuccessful applicants, but no evidence of such positive treatment effect for advanced researchers. Along a similar line, \citet{wang2018funding} compared the novelty of funded outputs from competitive projects and internal block projects, in the setting of Japan. They found on average a higher novelty for projects funded by competitive funds, but such privilege only holds for high status scientists and a negative relation to novelty is identified for low status researchers. 

These studies in essence seem to suggest that the selection process for grants exhibits a limited capacity in discerning and endorsing novel research proposals, especially from scholars with a more modest academic standing. This argument has been corroborated by numerous previous studies. \citet{Boudreau2016looking}, for example, noted that novel research proposals tend to garner lower ratings. \citet{ayoubi2021does} similarly found that novel scientists applying to a Swiss research funding program receive lower ratings and face fewer chances of being awarded.  

This funding bias against novel and risk research endeavors may be ascribed to the peer review process in grant selection that exhibits an undue fixation on feasibility and potential pitfalls and a propensity to overlook novel avenues underlying proposals \citep{azoulay2020scientific}. It may also be related to the effect of cognitive limit, which stems from intellectual distance that transcends reviewers' intellectual boundaries and leads them to potentially misconstrue proposals of a novel nature, thereby confounding a better grasp of the innovativeness in proposals. Some recent discussions have started to suggest possible interventions to mitigate biases against novelty and risk. In discussing challenges faced by researchers in getting funding support for mRNA technology, \citet{franzoni2021funding} delved into potential interventions that government agencies and universities could undertake to avoid such biases. \citet{franzoni2023uncertainty} discussed and clarified the concepts and meanings of risk and uncertainty in science and further suggested possible ways to incorporate risk into the peer review selection process of research proposals. 

The second contributing factor of the decline of novel science concerns about evaluations of researchers at institutions. The over-emphasis on productivity in existing evaluation exercises of researchers provides little incentives to them to pursue novel research endeavors, as an elevated risk profile in novel science indicates that a lower number of papers can be published, given positivity bias in current science publishing. \citet{hollingsworth2004institutionalizing} and \citet{heinze2020institutional} contended that the institutional environment where researchers are located, including a more stringent assessment system instituted by universities and research establishments, fails to provide a more risk-taking research climate, directly slowing down the pace of novel research exploration.

\subsection{Cross-country comparison of science and technology indicators}

Measuring and evaluating a country's status and standings in science and technology is a critical activity, as evaluation outcomes will have implications for science and technology policy-making. Many studies have presented cross-country comparative analyses of publications and citations \citep{narin1975national, may1997scientific, horta2007opening, King-nation-2004, XieKillewald2012}. These studies have indicated that the US has been the world center of science in the past century, accounting for a significant share of the world's research activities and output as well as impactful research. Recently, there has been attentions paid to the decline of US's dominance in science, accompanying by the rise of emerging countries, especially China. \citet{xie2014china} presented a comparative study on a number of aspects between the US and China, including science and engineering workforce and research output and impact. \citet{xie2019bigger} concluded that China's share in global science would be much higher when considering papers written in Chinese and by Chinese authors who are not affiliated with Chinese institutions. Studies focusing on individual fields also confirm the rise of China \citep{GUAN2007880}. Despite large in size, the debate that comes with it is whether contributions to scientific breakthroughs from China commensurate to its rise \citep{xie2014china}; China has produced a small percentage of top cited papers \citep{King-nation-2004} or papers appeared in reference lists of top cited papers \citep{bornmann2018geography}. Yet, in fields like artificial intelligence China belongs to prolific contributors \citep{OMeara2019will}. 

Another line of literature has further examined discipline structure and its implications in country performance. \citet{cimini2014scientific} built a bipartite network connecting countries and disciplines to obtain competitiveness of nations. Their analysis pointed out the nested nature of the bipartite network. A recent work, however, suggested that global science has moved from a nested structure to a modular structure \citep{miao2022latent}. Similarly, \cite{moya2013worldwide} perceived three clusters (the biomedical, basic science \& engineering, and agricultural clusters) and associated countries with the clusters. Such modular structure indicates that countries belong to different clusters of disciplines, implicating a division of ``labor'' in the global collaborative production of scientific knowledge by countries. Some studies have tried to explain disciplinary specializations of countries. For example, \citet{klavans2017research} devised the altruism and economic growth motivations behind research strategies of nations. \citet{krammer2017science} pointed out that by closely monitoring a nation's scientific exports and promptly adjusting its research strategies, it is possible to prevent a disconnect from economic objectives. 

Studies mentioned above have treated all papers equal and exclusively emphasized citation-based impact measures. Yet, as pointed out before, impact is an \emph{ex post} notion, and so far there has been very limited understanding of how nations perform in terms of \emph{ex ante} based productivity indicators. This work contributes to one such study that employs novelty and atypicality to examine how countries pursue creative research.

\section{Data and methods} \label{sec:data}

\subsection{Data}

This work aims to study the geography of creativity in science. We adopt the novelty and atypicality metrics for papers as operationalizations of creativity \citep{Uzzi-atypical-2013}. As the original proposals were implemented in the Web of Science (WoS) database, which is not publicly available, here we use a version of the MAG dataset retrieved in September 2020. Note that there is a major difference in terms of how reference data are given in the raw WoS and MAG databases. Specifically, in the WoS database, each cited reference of a document is given as a piece of text together with the corresponding identifier assigned by WoS (\emph{i.e.}, the accession number), so that even if the reference is not indexed in WoS, one can still perform further extractions to identify whether it is, for example, a preprint, a conference paper, or a book. By contrast, in the raw MAG data files, text representations of cited references are not provided. Instead, cited references have already been parsed and are listed as pairs of document identifiers indicating citing-cited relations. As such, during our data preparation stage, there are no other preprocessing steps involved related to the identifications of documents not indexed in MAG. 

From the MAG dataset, we build a paper citation network, $D$, which is used as the input for novelty calculations. To follow closely the procedure outlined in \citet{Uzzi-atypical-2013}, nodes in $D$ are the documents that are identified as journal articles by MAG, given that WoS indexes primarily journal papers. Documents that are excluded at this step are conference proceedings, book chapters, books, etc. Links in $D$ represent citation relationships between papers in the network. This means that cited references to journal papers are included, while references to non-journal documents are ignored. To comprehend the amount of retained references, we present several related statistics. First, there are $56,098,424$ papers in $D$, and they in total have $1,149,608,432$ references, among which $1,034,671,890$ (90\%) are to other papers in $D$. That is, 10\% of references are ignored, which remains stable for papers across publication years [Supporting Information (SI) Figure~S1A], suggesting a uniform processing by MAG. Second, to ascertain that papers with more references do not bias these statistics, we further focus on subsets of papers grouped by their reference count. SI Figure~S1B shows that the fractions of retained references are comparably high for different groups of papers, especially during the examined period between 1950 and 2019. 

Finally, to better understand the papers in $D$, we attempt to identify their languages based on multiple sources (see SI Section~S1 for details). We are able to do so for 81.3\% papers, and the results indicate that English accounts for the majority (at least 77.6\%) of the papers, whereas papers written in German (1.47\%), French (0.796\%), Spanish (0.414\%), and other languages are minorities (SI Table~S1).

\subsection{Classifying papers based on conventionality and novelty}

We implement the method presented in \citet{Uzzi-atypical-2013} to classify papers based on their conventionality and novelty scores. The method rests on the premises that journals are a viable representation of knowledge components, as each journal publishes a topically coherent set of papers, and that the extent of an article making conventional or novel combinations of existing knowledge components is captured by its referencing behaviors, specifically, how often it cites pairs of precedent works that are conventional or novel. Below we briefly describe the three major steps for the calculation of a paper's conventionality and novelty scores.
\begin{enumerate}
\item For each calendar year $t$, we obtain the observed journal co-occurrence matrix $O^t$ where the entry $o_{ij}^t$ counts the total number of times journals $i$ and $j$ co-occur in the reference lists of papers published in year $t$.
\item The second step compares the observed frequency $o_{ij}^t$ with expected frequency by chance. To elaborate, we randomize the entire citation network $D$ formed by all the papers in our sample for 10 times. Each randomization, $r$, creates reshuffled lists of references, from which we can get a randomized journal co-occurrence matrix $\tilde{O}^{t, (r)}$. We then calculate the journal pair $z$-score matrix $Z^t$, where the entry $z_{ij}^t$ is defined as
\begin{equation}
z_{ij}^t = \frac{o_{ij}^t - \text{mean}\left[ \tilde{o}_{ij}^{t, (r)} \right]}{\text{std}\left[ \tilde{o}_{ij}^{t, (r)} \right]} \, .
\end{equation}
If a journal pair's $z$-score is less than zero, it indicates that the pair is observed less frequently than expected by chance, therefore more novel, whereas journal pairs with $z$-score above zero indicate that they are observed more frequently than expected, hence more conventional.
\item For a paper published in $t$, we get a distribution of $z$-scores based on its referenced journal pairs. We take the median $z$-score as a paper's conventionality score and 10-th percentile $z$-score as its novelty score.
\end{enumerate}

Given the two sets of scores, we proceed to categorize papers along two dimensions. First, a paper is considered as highly novel if its novelty score is less than zero. As this classification is not dependent on publication year, we can study novelty at a yearly basis. Second, a paper is classified as highly conventional if its conventionality score is above the median conventionality score of all papers in a decade. Note that by construction, half of papers published in a decade are categorized as highly conventional. A paper is deemed atypical if it is both highly novel and highly conventional. As the categorization of conventionality is made for every decade, our analyses about atypicality are also at a decade basis. We perform this categorization for $37,989,504$ papers published between 1950 and 2019. Finally, to facilitate further research on creativity in science, we have made publicly available conventionality and novelty scores and corresponding classifications of all MAG papers in our sample at \url{https://doi.org/10.5281/zenodo.15250119}.

\subsection{Extracting countries from affiliations}

Having classified MAG papers, we can study geographic origins of articles falling into the highly novel and the atypical categories. In so doing, we rely on author affiliations to geolocate papers. Papers in our sample in total have $158,238,120$ authorships (one author in one paper) and $162,845,097$ affiliations, since one author can be affiliated with more than one organization. To obtain countries where these affiliations are located, we take advantage of affiliation related information provided by MAG and obtain country level locations using affiliation identifiers, which accounts for $109,984,270$ (67.5\%) affiliations. For $27,506,024$ (16.9\%) affiliations, there are textual information but no identifiers associated with them. We therefore extract country directly from affiliation text, and we are able to do so for $77.7\%$ of this type of affiliations. For the remaining $25,354,670$ (15.6\%) affiliations, there are no text or identifiers.

We present in SI Figure~S2 fractions of papers published over time for which we are able to extract country. It indicates that the fraction of papers where no authors' country location can be extracted has been decreasing drastically over time. Starting from the 1990s, for more than 70\% of papers, all of their authors have available country information, and the percentage is even higher if we consider papers where at least half of their authors have country information. Based on these statistics, when studying geography of novelty and atypicality, we focus on the $26,110,499$ unique papers published from 1990 and onward where at least half of authors have available affiliation. SI Table~S1 provides their language decomposition, showing that the vast majority (more than 90\%) of this subset of papers are in English and suggesting that our results are not driven by non-English papers. 

After extracting countries from affiliations, we count the number of papers published by a country. We follow the fractional counting procedure, which assigns each paper one unit to be equally spitted among its coauthors, and the amount of unit of a coauthor is further shared among the affiliations (s)he belongs to. In this way, papers with more coauthors will not weight more, and there are $25,461,110.87$ papers with country information. 

\subsection{Quantifying the tendency to produce novel and atypical research}

After geolocating papers, we quantify a country's tendency to produce novel and atypical papers during a period of time. We use the case of novel papers to illustrate the quantification. For each country, we calculate the ratio, $r_N$, between its share in novel papers published in that period and its share in all papers published in the same period. To elaborate, let $M$ be the total number of papers published during a period of time, among which $n$ papers are classified as novel ones; for a particular country, it publishes $C$ papers, among which $j$ papers are novel. Then, we define $r_N = \frac{j}{n} \bigg/ \frac{C}{M}$. A country with $r_N > 1$ indicates that it produces novel papers disproportionately, \emph{i.e.}, it is over-represented in the set of novel papers, whereas $r_N < 1$ means it is under-represented in novel papers, thereby providing a quantification of the country's tendency to generate novel publications relative to other countries. The tendency to publish atypical papers, $r_A$, is similarly defined.

We further assess statistical significance of $r_N$ and $r_A$ using the hypergeometric test. Both measures take the total number of papers published by a country into consideration, because a country with more papers would have more novel/atypical papers by chance. The significance test therefore is to examine whether the deviation of $r_N$ ($r_A$) from one can be explained by size (\emph{i.e.}, the number of papers) alone. If no additional factors other than size play a role, then the expected number of novel papers, $k$, is a random variable following hypergeometric distribution, with probability mass function given by $\text{Pr}\left(X = k\right) = {\binom{n}{k}}{\binom{M-n}{C-k}} \bigg/ {\binom{M}{C}}$. Therefore, if $r_N > 1$, we perform the hypergeometric test for over-representation, and the $p$-value is the probability of having at least $j$ papers from that country that are novel: $p = 1 - \sum_{k=0}^{j-1} \text{Pr}\left(X = k\right)$. If $r_N < 1$, we perform the hypergeometric test for under-representation, and the $p$-value is the probability of having at most $j$ papers from that country that are novel: $p = \sum_{k=0}^{j} \text{Pr}\left(X = k\right)$.

\section{Results} \label{sec:res}

\subsection{Reproducing Uzzi et al. (2013)}

\begin{figure*}[t!]
\centering
\includegraphics[trim=0 2mm 0 0, width=\textwidth]{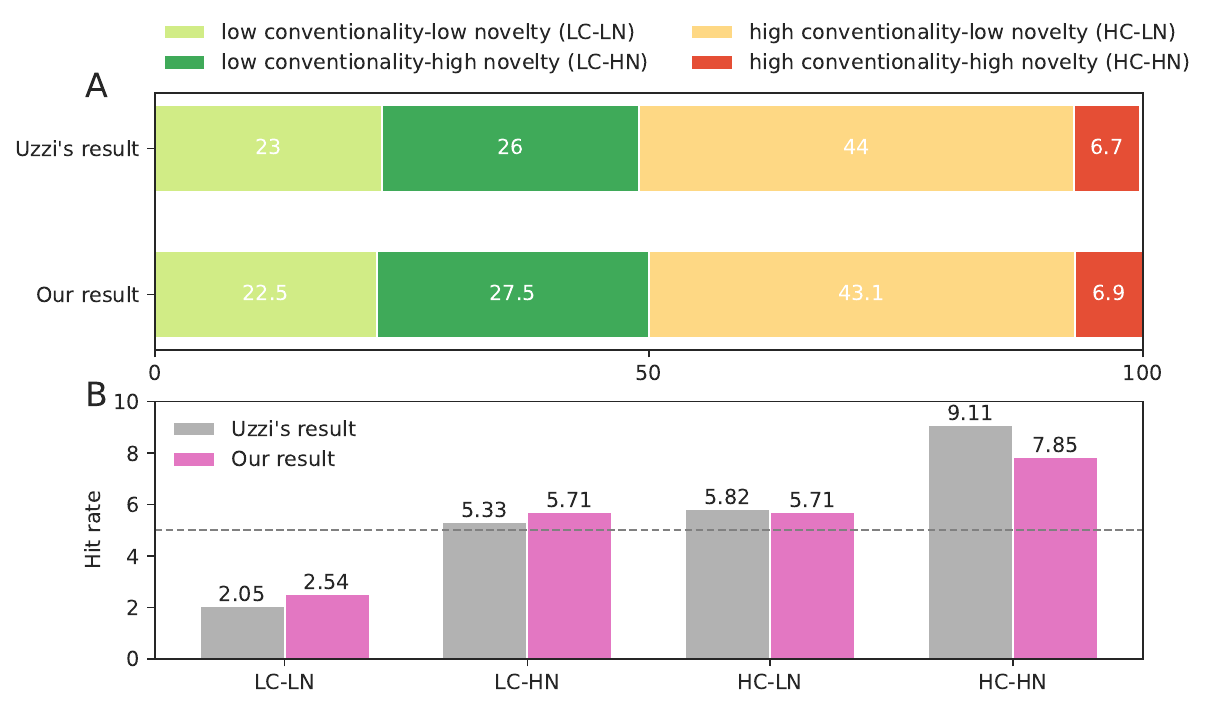}
\caption{Reproducing the relationship between atypical science and scientific impact, as reported in \citet{Uzzi-atypical-2013}. The figure considers all papers published in the 1990s. (A) The compositions of the four groups of papers based on our implementation, which uses MAG, and based on \citet{Uzzi-atypical-2013}, which used WoS, indicating that each group's relative share remains almost identical despite different datasets are used. (B) Hit rates of the four groups of papers in MAG and WoS, showing that the HC-HN group has the highest hit rate. Hit rate is defined as the percentage of papers in a group whose citations are ranked into the top 5\% of all papers. The dashed horizontal line marks the baseline hit rate (5\%).}
\label{fig:atypical-impact}
\end{figure*}

To validate our implementation of the classifications of papers based on their conventionality and novelty as well as to check whether the findings from \citet{Uzzi-atypical-2013} are robust to different bibliographic datasets, we set out to examine whether some key results in \citet{Uzzi-atypical-2013} can be reproduced. First, focusing on papers published in the 1990s, Figure~\ref{fig:atypical-impact}A reports the compositions of the four groups of papers based on their conventionality and novelty, using the MAG dataset (our implementation) and the WoS dataset (results from \citet{Uzzi-atypical-2013}). It shows that the relative share of each group of papers remains almost identical in the two datasets and that atypical papers---those classified as highly conventional and highly novel (``HC-HN'')---account for the smallest portion of the scientific literature. Second, \citet{Uzzi-atypical-2013} found that there is a ``powerful'' relationship between atypicality and scientific impact, indicating that atypical papers are more (twice as) likely to be highly cited articles. This relationship is reproduced in Figure~\ref{fig:atypical-impact}B, demonstrating that among the four groups of papers, the atypical one has the highest hit rate, defined as the percentage of hit papers---papers whose 8-year citations are ranked into the top 5\% of all papers---in the group. The exact magnitude of hit rate differs across the two datasets; for the HC-HN group, the hit rate is 7.85 using MAG, while it is 9.11 using WoS. The difference in hit rates might due to several reasons. For example, the sample of MAG papers is much larger than the WoS sample, which might lead to a different set of high impact papers, which might be less likely to belong to the HC-HN category.

\begin{figure*}[t!]
\centering
\includegraphics[trim=0 2mm 0 0, width=\textwidth]{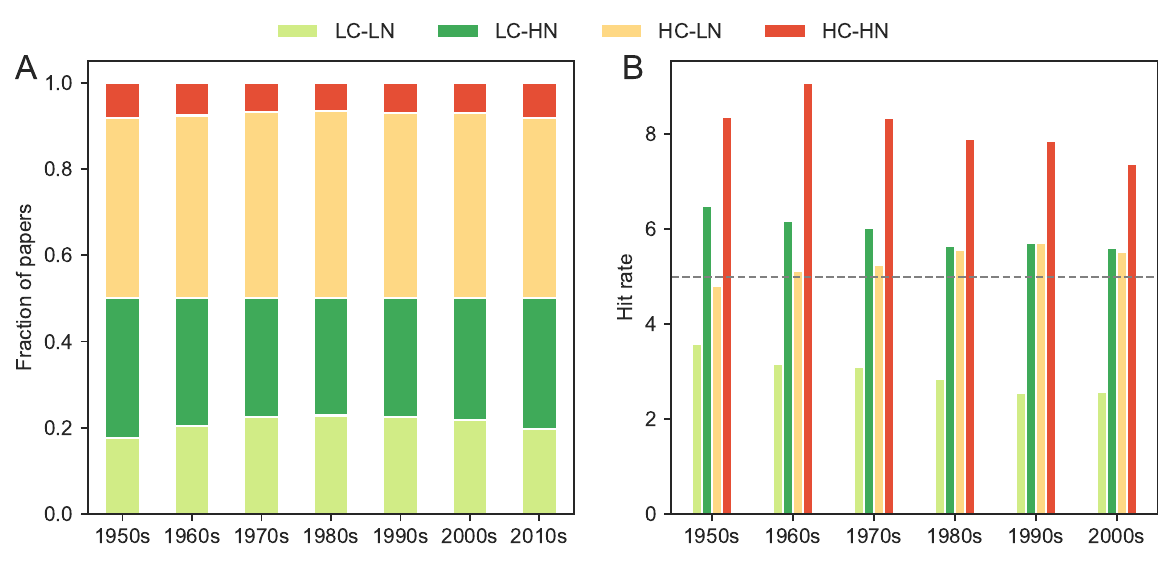}
\caption{The persistent relationship between atypicality and scientific impact across decades. (A) Compositions of the four categories of MAG papers published in each decade, suggesting that atypical papers are the least common in science. (B) Hit rates of the four groups of papers by decade, indicating that atypical papers are most likely to be hits.}
\label{fig:hit-decade}
\end{figure*}

\citet{Uzzi-atypical-2013} further showed that the superior citation performance of atypical papers is a universal phenomenon that can be observed for papers published in different decades. We therefore examine this by repeating our analyses for MAG papers by decade. We find that the results above are generalizable to other decades. The group of atypical papers continues to be the smallest category across the seven decades from the 1950s to the 2010s (Figure~\ref{fig:hit-decade}A), but consistently has the largest probability to have hit works across the six decades (Figure~\ref{fig:hit-decade}B; papers published in the 2010s are excluded, as citation-window is not long enough). However, we note that the hit rate of atypical papers has been decreasing since 1960s, a trend that has also been observed in \citet{Uzzi-atypical-2013}. In the 1960s, the hit rate was 9.08, which decreased to 7.37 in the 2000s. Such a decrease suggests a diminishing power of novel-conventional knowledge combination in contributing to the production of high impact science, and might due to the connections to canonization of preexisting knowledge playing an increasingly important role in getting studies ranked into the highly cited group \citep{chu2021slowed}. 

To summarize, our reproducing results lend strong support for the validity of our implementation of paper classifications as well as for the robustness of the association between atypicality and scientific impact.

\subsection{Countries producing novel research}

Having validated our implementation of conventionality and novelty classifications, next we study geographic origins of high novelty papers and atypical papers at the country level. We focus on the list of 38 selected countries/regions, which in total account for over 95\% of all papers. This list is the union of the top 30 largest producers in the entire 30-year period from 1990 to 2019 and the top 30 in each year. Table~\ref{tab:country-overall} reports various statistics about all papers and novel and atypical ones for these countries/regions. The US is the most prolific country, contributing 27.23\% of the world's scientific output, followed by China (8.69\%), UK (6.54\%), and Japan (5.7\%). When looking at the set of high novelty papers, 27.63\% came from the US, making it over-represented in the novel paper set compared with its share in overall papers ($r_N = 1.01$, $p < 0.001$). China, the second largest producer in science, has a 9.01\% portion in novel papers, indicating that the extent of its over-representation ($r_N = 1.04$, $p < 0.001$) is greater than that of the US, which means that during this period, China is more likely to publish novel papers than the US. UK, on the other hand, is under-represented in novel papers ($r_N = 0.96$). In general, we observe that $r_N$ is narrowly distributed around one, suggesting that for the examined countries/regions, their under-/over-representations in novel papers, though statistically significant (except Norway), are bibliometrically small in effect size; that is, their shares in novel papers are relatively similar to shares in research output overall. The exception is Russia, which is under-represented in the set of novel papers, with its $r_N$ (0.78) appearing to be away from the rest. Following Russia, Hong Kong, Spain, Singapore, South Africa, and Greece have the smallest $r_N$, whereas Egypt, Brazil, Malaysia, Denmark, and Sweden are the five countries with the largest $r_N$.

\begin{table*}[t!]
\centering
\caption{Statistics about the production of science and novel and atypical science in 1990--2019 for a list of 38 countries/regions. SI Tables S3--S5 provide the same set of statistics for each decade separately.}
\label{tab:country-overall}
\begin{tabular}{l |  rr | rrcc | rrcc}
\toprule
 & \multicolumn{2}{c|}{All papers} & \multicolumn{4}{c|}{Novel papers} & \multicolumn{4}{c}{Atypical papers} \\
\multicolumn{1}{l|}{Country/region} & \multicolumn{1}{c}{Count} & \multicolumn{1}{c|}{\%} & \multicolumn{1}{c}{Count} & \multicolumn{1}{c}{\%} & \multicolumn{1}{c}{$r_N$} & \multicolumn{1}{c|}{$p$-val} & \multicolumn{1}{c}{Count} & \multicolumn{1}{c}{\%} & \multicolumn{1}{c}{$r_A$} & \multicolumn{1}{c}{$p$-val} \\
\midrule
US & 6932688.28 & 27.23 & 2662828.96 & 27.63 & 1.01 & *** & 594625.35 & 29.13 & 1.07 & *** \\
China & 2213413.12 & 8.69 & 868340.98 & 9.01 & 1.04 & *** & 232482.10 & 11.39 & 1.31 & *** \\
UK & 1664853.72 & 6.54 & 606934.47 & 6.30 & 0.96 & *** & 114801.13 & 5.62 & 0.86 & *** \\
Japan & 1451048.40 & 5.70 & 544748.13 & 5.65 & 0.99 & *** & 123667.72 & 6.06 & 1.06 & *** \\
Germany & 1221638.22 & 4.80 & 452645.70 & 4.70 & 0.98 & *** & 106634.37 & 5.22 & 1.09 & *** \\
France & 901661.61 & 3.54 & 327470.48 & 3.40 & 0.96 & *** & 75697.04 & 3.71 & 1.05 & *** \\
Canada & 839601.40 & 3.30 & 309402.54 & 3.21 & 0.97 & *** & 62503.54 & 3.06 & 0.93 & *** \\
India & 762481.88 & 2.99 & 302340.29 & 3.14 & 1.05 & *** & 60573.79 & 2.97 & 0.99 & ** \\
Italy & 762474.39 & 2.99 & 290630.87 & 3.02 & 1.01 & *** & 58694.41 & 2.88 & 0.96 & *** \\
Australia & 710618.09 & 2.79 & 259063.42 & 2.69 & 0.96 & *** & 46171.58 & 2.26 & 0.81 & *** \\
Spain & 660647.15 & 2.59 & 225505.34 & 2.34 & 0.90 & *** & 50286.61 & 2.46 & 0.95 & *** \\
Korea & 618651.29 & 2.43 & 239476.70 & 2.49 & 1.02 & *** & 51518.05 & 2.52 & 1.04 & *** \\
Brazil & 539665.02 & 2.12 & 223343.25 & 2.32 & 1.09 & *** & 31767.35 & 1.56 & 0.73 & *** \\
Netherlands & 441287.02 & 1.73 & 176225.68 & 1.83 & 1.06 & *** & 36258.26 & 1.78 & 1.02 & *** \\
Russia & 418100.28 & 1.64 & 122911.77 & 1.28 & 0.78 & *** & 23386.63 & 1.15 & 0.70 & *** \\
Taiwan & 329353.26 & 1.29 & 121648.50 & 1.26 & 0.98 & *** & 26191.76 & 1.28 & 0.99 &  \\
Iran & 320312.16 & 1.26 & 130748.54 & 1.36 & 1.08 & *** & 23592.15 & 1.16 & 0.92 & *** \\
Sweden & 313696.19 & 1.23 & 128051.62 & 1.33 & 1.08 & *** & 23885.30 & 1.17 & 0.95 & *** \\
Turkey & 269882.18 & 1.06 & 99835.65 & 1.04 & 0.98 & *** & 15826.35 & 0.78 & 0.73 & *** \\
Switzerland & 256439.11 & 1.01 & 98684.37 & 1.02 & 1.02 & *** & 23920.32 & 1.17 & 1.16 & *** \\
Poland & 248930.63 & 0.98 & 92716.46 & 0.96 & 0.98 & *** & 15241.02 & 0.75 & 0.76 & *** \\
Belgium & 219250.53 & 0.86 & 81978.55 & 0.85 & 0.99 & *** & 17889.82 & 0.88 & 1.02 & ** \\
Israel & 187230.04 & 0.74 & 69930.19 & 0.73 & 0.99 & *** & 13866.88 & 0.68 & 0.92 & *** \\
Denmark & 177004.82 & 0.70 & 72530.53 & 0.75 & 1.08 & *** & 14248.06 & 0.70 & 1.00 &  \\
Mexico & 160385.81 & 0.63 & 63242.60 & 0.66 & 1.04 & *** & 11082.42 & 0.54 & 0.86 & *** \\
Finland & 151982.23 & 0.60 & 59462.23 & 0.62 & 1.03 & *** & 11440.83 & 0.56 & 0.94 & *** \\
South Africa & 143140.86 & 0.56 & 50148.80 & 0.52 & 0.93 & *** & 7137.49 & 0.35 & 0.62 & *** \\
Austria & 140691.05 & 0.55 & 54856.43 & 0.57 & 1.03 & *** & 11344.02 & 0.56 & 1.01 &  \\
Norway & 137786.36 & 0.54 & 52233.82 & 0.54 & 1.00 &  & 9146.08 & 0.45 & 0.83 & *** \\
Greece & 131670.84 & 0.52 & 46927.71 & 0.49 & 0.94 & *** & 8997.90 & 0.44 & 0.85 & *** \\
Hong Kong & 130085.63 & 0.51 & 43582.63 & 0.45 & 0.89 & *** & 10112.08 & 0.50 & 0.97 & *** \\
Portugal & 125977.19 & 0.49 & 48520.67 & 0.50 & 1.02 & *** & 9008.78 & 0.44 & 0.89 & *** \\
New Zealand & 117689.24 & 0.46 & 42706.44 & 0.44 & 0.96 & *** & 7889.51 & 0.39 & 0.84 & *** \\
Singapore & 114029.09 & 0.45 & 39756.86 & 0.41 & 0.92 & *** & 10243.08 & 0.50 & 1.12 & *** \\
Malaysia & 104243.55 & 0.41 & 42873.82 & 0.44 & 1.09 & *** & 6399.50 & 0.31 & 0.77 & *** \\
Czechia & 99740.13 & 0.39 & 36201.93 & 0.38 & 0.96 & *** & 6971.54 & 0.34 & 0.87 & *** \\
Egypt & 97240.11 & 0.38 & 42162.94 & 0.44 & 1.15 & *** & 6004.28 & 0.29 & 0.77 & *** \\
Hungary & 77023.84 & 0.30 & 28883.09 & 0.30 & 0.99 & * & 5429.36 & 0.27 & 0.88 & *** \\
\midrule
\multicolumn{1}{r|}{\emph{Total}} & 24192614.68 & 95.02 & 9159552.95 & 95.05 & -- & -- & 1964936.47 & 96.26 & -- & -- \\
\bottomrule
\multicolumn{11}{l}{\footnotesize \sym{*} \(p<0.05\), \sym{**} \(p<0.01\), \sym{***} \(p<0.001\)}\\
\end{tabular}
\end{table*}

\begin{figure*}[t!]
\centering
\includegraphics[trim=0 2mm 0 0, width=\textwidth]{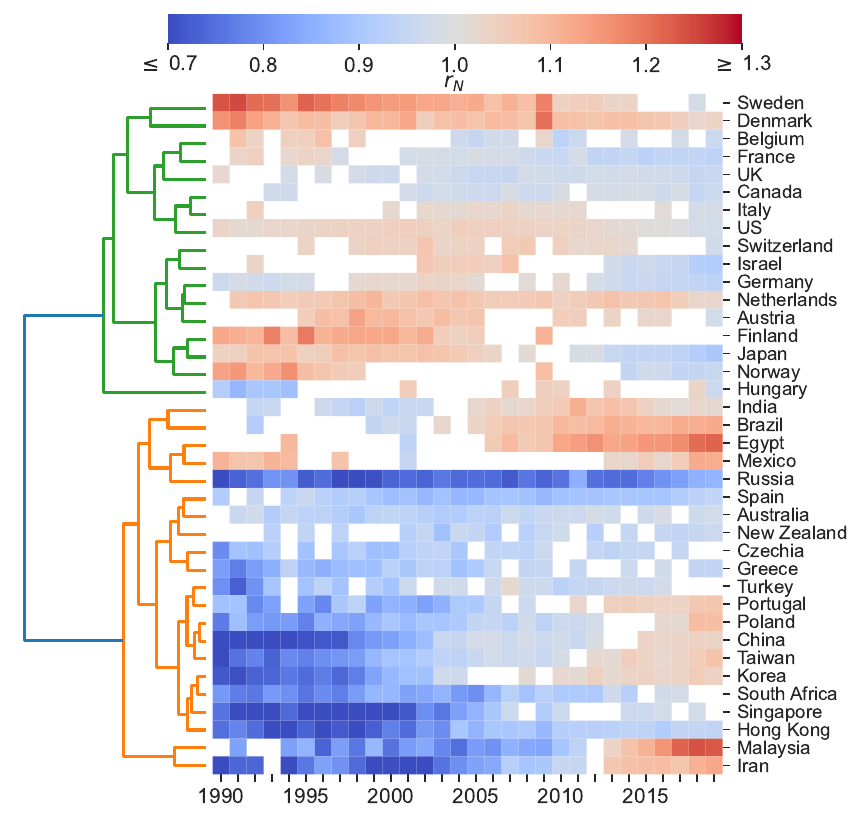}
\caption{The tendency $r_N$ of the production of novel papers by country/region during 1990--2019. Color encodes $r_N$.}
\label{fig:country-novelty}
\end{figure*}

These overall statistics mask large variations of how the tendency of publishing novel research changes temporarily. We therefore unfold $r_N$ over time and characterize its temporal patterns. We recalculate $r_N$ for each of the 38 countries/regions in each year from 1990 to 2019, and Figure~\ref{fig:country-novelty} visualizes this $38 \times 30$ matrix of $r_N$, where cases that are lack of statistical significance at the 0.05 level are not shown. To uncover groups of countries that exhibit similar temporal changes, we perform hierarchical clustering of these countries based on their time series of $r_N$, with the clustering results displayed as the dendrogram on the left. We identify two large clusters. The first one consists of 17 countries, which are Sweden, Denmark, Belgium, France, Canada, UK, Italy, Switzerland, US, Israel, Germany, Austria, Netherlands, Finland, Norway, Japan, and Hungary. These are mostly western countries that are traditional players in science publishing. On average, they are more likely to generate novel research in the 1990s, but lose this tendency over time. Notably, within this cluster, Sweden, Denmark, US, and Netherlands have been over-represented in novel papers for an extended period of time. The second cluster contains 21 countries/regions, which are India, Brazil, Egypt, Mexico, Russia, Spain, New Zealand, Australia, Czechia, Turkey, Greece, Poland, Portugal, China, Korea, Taiwan, South Africa, Singapore, Hong Kong, Iran, Malaysia. They can be further decomposed into two subclusters: (1) emerging regions that have made a recent transition from being under-represented to over-represented in novel papers, including Poland, Portugal, China, Korea, Taiwan, Iran, and Malaysia; and (2) the rest that exhibits no noticeable improvement in their representation in novel science.

\subsection{Countries producing atypical research}

Similarly, we study countries' tendency, $r_A$, to produce atypical papers. Although the classification of atypical papers is made for each decade due to its dependence on the conventionality classification, which is based on papers published in a decade, we nevertheless examine $r_A$ for the entire 30-year period to gain an overall understanding. The rightmost four columns in Table~\ref{tab:country-overall} show relevant statistics about atypical papers by country/region. China stands out as the country that is most over-represented ($r_A = 1.31$), followed by Switzerland ($1.16$), Singapore ($1.12$), and Germany ($1.09$). On the other end of the spectrum, South Africa ($0.62$), Russia ($0.7$), and Turkey ($0.73$) are the three countries that are most under-represented in atypical papers. What is also interesting is countries that display a sharp contrast between representations in novel and atypical papers. Brazil, for example, is the most novel country ($r_N = 1.09$), yet the fourth least atypical one ($r_A = 0.73$). This means that Brazilian papers have introduced novel combinations of existing knowledge components without making enough conventional ones. Conversely, Singapore is fourth least novel country ($r_N = 0.92$), but the third most atypical one.

\begin{figure*}[t!]
\centering
\includegraphics[trim=0 2mm 0 0, width=.95\textwidth]{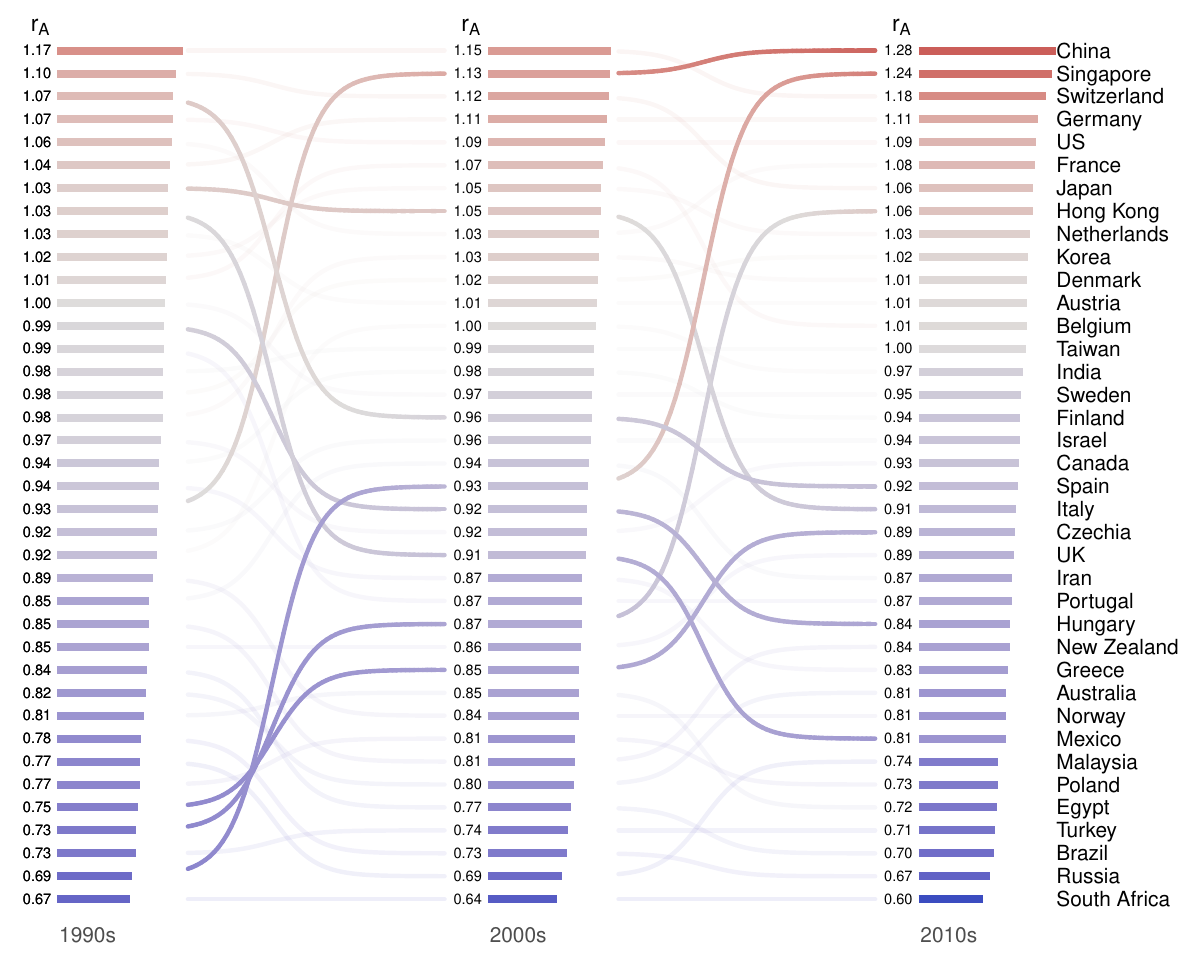}
\caption{Ranking of countries/regions based on their $r_A$, the tendency to produce atypical papers, in the three decades. Ranking is within the 38 selected places, and color encodes $r_A$.}
\label{fig:country-atypicality}
\end{figure*}

We further study how $r_A$ changes from 1990s to 2010s. We recalculate $r_A$ for each decade and then rank countries/regions based on this metric. Figure~\ref{fig:country-atypicality} presents the ranking results, with countries/regions with drastic ranking changes highlighted. In the 1990s, Switzerland, Japan, and Spain are the three countries that were most likely to have atypical papers. While the first two continued to rank high in the following two decades, Spain quickly lost its leadership role and became under-represented in atypical research. Other countries with deceasing rankings include Mexico, Hungary, and Portugal. Rapid increases in the ranking are apparent for China, Singapore, and Hong Kong. The rankings for the US remain relatively stable.

\begin{figure*}[t!]
\centering
\includegraphics[trim=0 4mm 0 0, width=\textwidth]{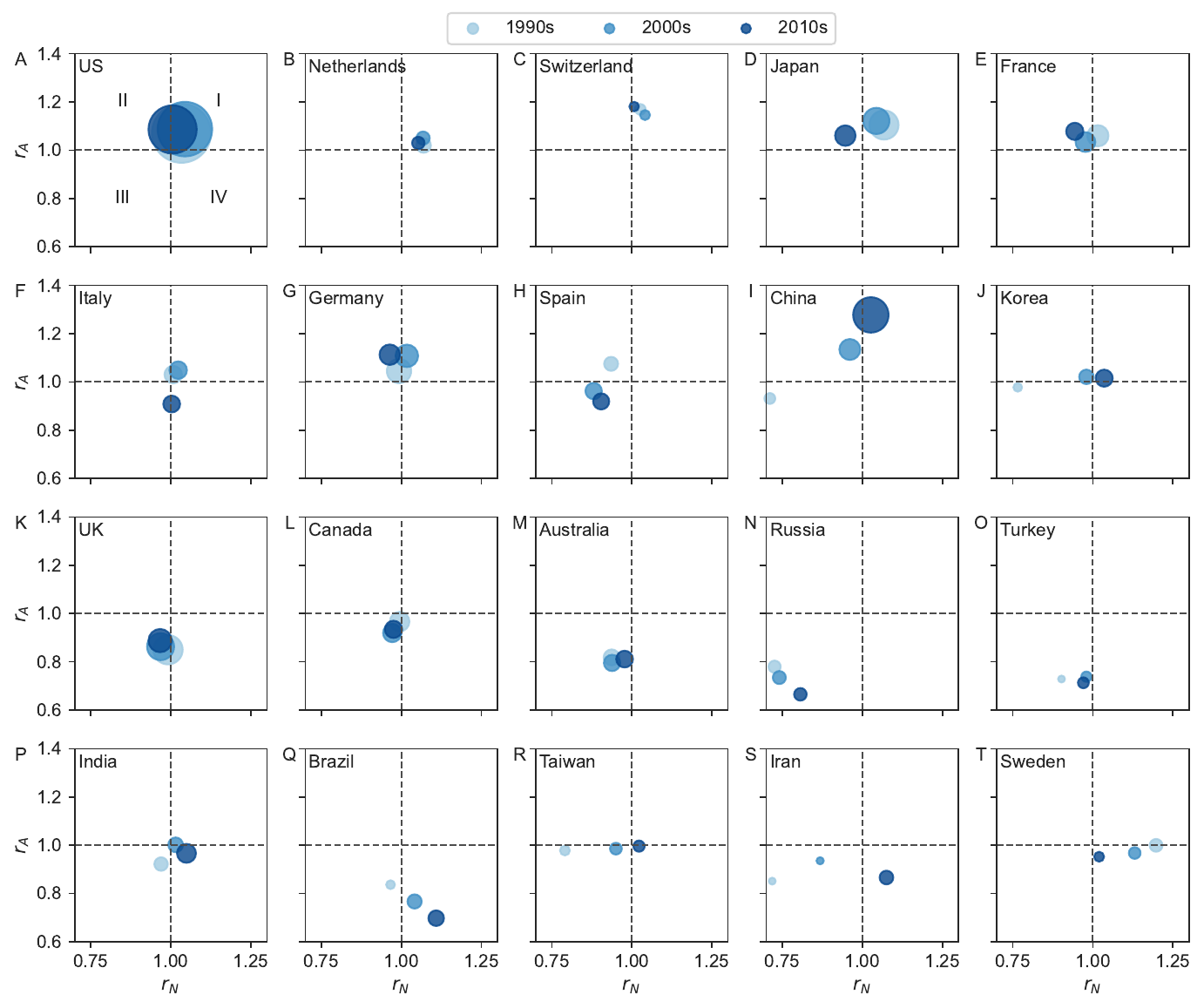}
\caption{Comparison of the tendency to generate novel and atypical science. Color represents decade, and scatter size indicates relative share among all papers published in the decade.}
\label{fig:novelty-atypicality}
\end{figure*}

Finally, we group countries based on how their $r_N$ and $r_A$ have changed during the last three decades. We partition the $r_N$--$r_A$ space into four quadrants based on whether $r_N$ and $r_A$ are above or below one, whereby quadrant I corresponds to $r_N > 1$ and $r_A > 1$ (Figure~\ref{fig:novelty-atypicality}A). We find that countries exhibit different transition paths (Figure~\ref{fig:novelty-atypicality}) and describe in detail below.
\begin{enumerate}
    \item Starting from quadrant I in the 1990s (Figures~\ref{fig:novelty-atypicality}A--F): (1) US, the Netherlands, and Switzerland have stayed there in the 2000s and 2010s, keeping up their ability in generating novel and atypical research (Figures~\ref{fig:novelty-atypicality}A--C); (2) Japan and France have moved to quadrant II, losing their representation in novel papers (Figures~\ref{fig:novelty-atypicality}D--E); and (3) Italy has transitioned to quadrant IV, being under-represented in atypical science (Figure~\ref{fig:novelty-atypicality}F). 
    \item Germany and Spain are the two countries that were located in quadrant II in the 1990s. While the former has stayed in quadrant II (Figure~\ref{fig:novelty-atypicality}G), the latter has become less likely to produce novel and atypical research (quadrant III; Figure~\ref{fig:novelty-atypicality}H). 
    \item There are a number of countries/regions that under-performed in producing novel and atypical science in the 1990s (quadrant III). China and Korea are the two countries that have successfully moved to quadrant I, gaining their statuses of being over-represented in both novel and atypical science. However, their exact locations in the quadrant are rather different; Korea is situated close to the (1,1) point (Figure~\ref{fig:novelty-atypicality}J), indicating that its over-representation is bibliometrically marginal, whereas China has moved away from the (1,1) point (Figure~\ref{fig:novelty-atypicality}I). UK, Canada, Australia, Russia, and Turkey have not changed qualitatively in terms of their under-representation statuses during the three decades, remaining in quadrant III (Figures~\ref{fig:novelty-atypicality}K--O). India, Brazil, Taiwan, and Iran have proceeded to quadrant IV, showing improvement in novel science but not atypical science (Figures~\ref{fig:novelty-atypicality}P--S).
    \item Sweden continues to be over-represented in novel but not atypical science for the three decades (Figure~\ref{fig:novelty-atypicality}T).
\end{enumerate}

\subsection{Discipline and international collaboration effects}

\begin{figure*}[t!]
\centering
\includegraphics[trim=2mm 8mm 5mm 0mm, width=\textwidth]{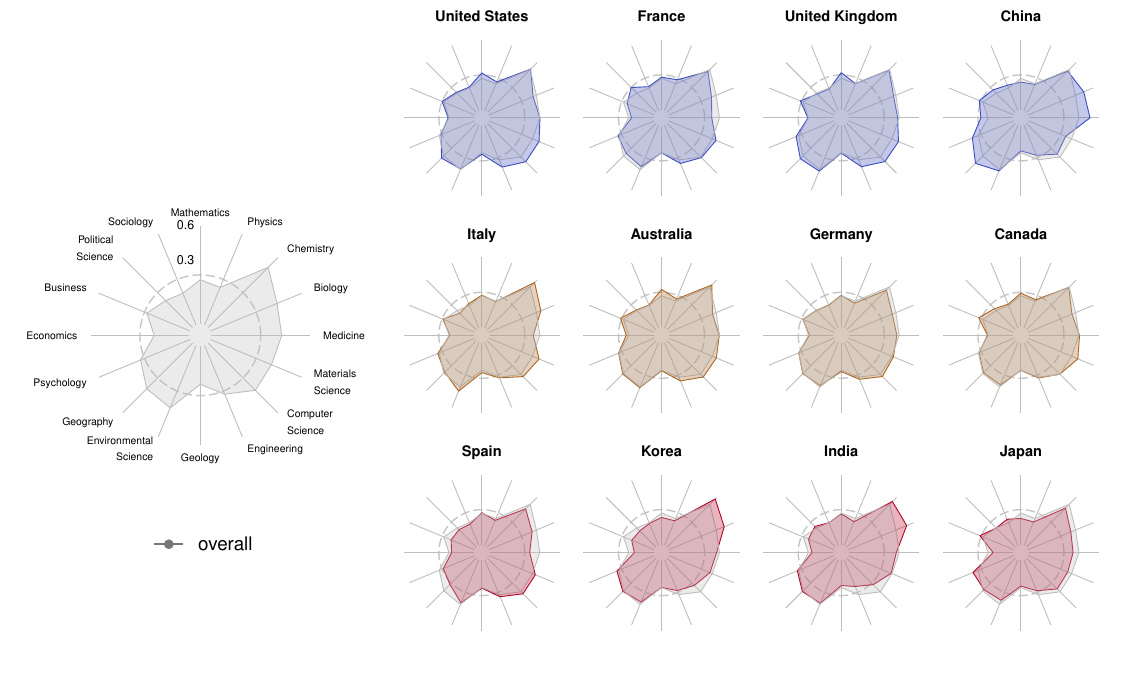}
\caption{Fraction of papers in the 2010s that are novel by country-discipline combination. The grey lines, which are duplicated across panels, correspond to overall fraction by discipline without considering country. We focus on the 16 disciplines that are located at the highest level in the field of study hierarchy generated by MAG and ignore philosophy, history, and art, due to a small number of papers.}
\label{fig:novelty-field-rate}
\end{figure*}

\begin{figure*}[t!]
\centering
\includegraphics[trim=2mm 8mm 5mm 0mm, width=\textwidth]{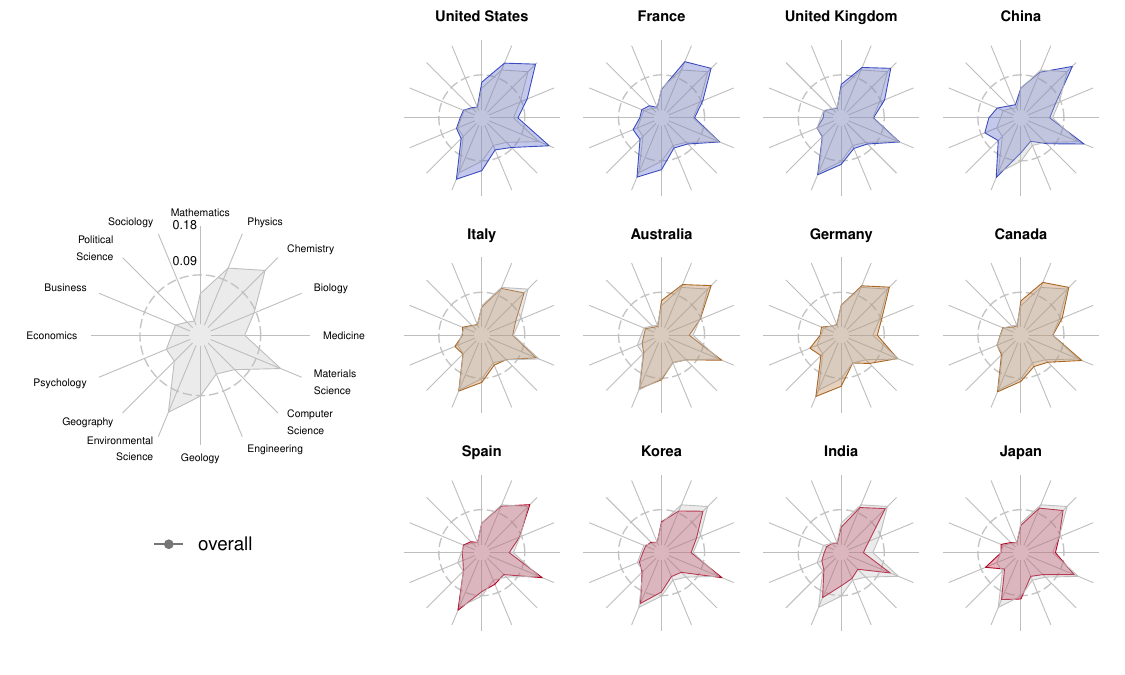}
\caption{Fraction of papers in the 2010s that are atypical by country-discipline combination.}
\label{fig:atypicality-field-rate}
\end{figure*}

Our analyses thus far have revealed that countries have varying extent of tendency in producing novel and atypical science. Such tendency, however, may be confounded by discipline effects, as countries may exhibit specialties in certain disciplines and the share of novel and atypical research may be dependent on disciplines. We therefore seek to understand how countries perform in different disciplines and to what extent disciplinary specialization can explain the differences in the production of novel and atypical research. We start by examining rates of novel and atypical papers in the 2010s by discipline, showing as the grey curves in Figures~\ref{fig:novelty-field-rate}--\ref{fig:atypicality-field-rate}, respectively. We see that there are indeed large discipline-level variations. Chemistry papers are most likely to be novel as well as atypical; about 52\% of chemistry papers are novel and 15\% atypical. This superiority is consistent with the argument that chemistry is ``the central science'', connecting physical science with life science, medicine, engineering, etc. \citep{balaban2006chemistry, szell2018nobel, peng2021neural}. A bridging role of chemistry suggests an expanded space for possible recombination, making it more likely to introduce novel one. Papers in social science fields, like sociology, political science, business, and economics, are relatively less likely to be highly novel or atypical, possibly reflecting limited knowledge ingredients with which social science papers are able to recombine.

Delving into countries/regions, each panel in Figures~\ref{fig:novelty-field-rate}--\ref{fig:atypicality-field-rate} focuses on one country and plots the fraction of novel and atypical papers in different country-discipline groups. To ease of comparison, we also duplicate the overall case (without considering country) in each panel. Therefore, for a particular country-discipline group, a fraction larger than the baseline indicates that, \emph{conditional on the discipline}, the country is over-represented in novel/atypical papers (\emph{i.e.}, $r_N^d$ or $r_A^d > 1$ for discipline $d$). It can be seen that for novel science (Figure~\ref{fig:novelty-field-rate}), the US and UK have similar profiles, performing better than the global average for mathematics, materials science, computer science, and engineering. China produces more novel papers in biology, medicine, geography, psychology, economics. Spain and Japan perform worse than global average for a number of fields, including physics, chemistry, and medicine. For atypical science (Figure~\ref{fig:atypicality-field-rate}),  the US over-produces in all the disciplines, whereas China's over-production of atypical science is limited to a few disciplines, especially the most prolific ones (Materials Science and Chemistry) that drive the overall performance. UK excels in most disciplines except Sociology and Economics, whereas Spain, Korea, India, and Japan generate atypical science at a rate lower than global average. To make the discipline effect more apparent, SI Tables~S7--S34 report detailed discipline specific performances of novel and atypical paper production by these focused countries. 

Finally, we examine the effect of international collaboration on novel and typical science. We first group papers based on whether they involve international collaboration and then recalculate the $r_N$/$r_A$ measures separately for the two groups. The results indicate that among the most prolific countries in the 2010s, US, China, and India are over-represented in novel papers for both groups, whereas UK, Japan, Germany, France, and Canada are under-represented (SI Table~S35). For atypical papers, US, China, Japan, Germany, and France over-produced, whereas UK, Italy, and Australia under-produced (SI Table~S36).

\section{Discussion} \label{sec:dis}

\subsection{Main findings}

The main purpose of this work was to examine the geography of the production of novel and atypical research at the country level. There has been increasing attention to novelty in science, due in part to its high-risk and high-return features. While existing literature has largely devoted to measuring novelty and examining its relationship with recognition, it remains unknown about geographic origins of novel science. 

To answer this question, we have implemented and tested the conventionality and novelty classifications of MAG papers, allowing us to reproduce \citet{Uzzi-atypical-2013} at a large and systematic scale. We have further introduced an indicator that quantifies a country's representativeness in novel and atypical papers compared with its overall scientific output, finding that differences in propensity to generate novel and atypical science persist among countries. This is the case even among developed countries that are large producers in science. Currently, among the top ten largest producers, US, China, India, and Italy are over-represented in novel papers, whereas UK, Japan, Germany, France, Canada, and Australia are under-represented; for atypical science, US, China, Japan, Germany, and France are over-represented, whereas UK, Canada, India, Italy, and Australia are under-represented. We further find that there is rich dynamics in countries' representativeness in novel science such that we are able to cluster them, with different groups of countries identified. One cluster is composed by a number of countries that are emerging as leading producers that disproportionately generate more novel science, including Brazil, China, and Korea. Looking at atypical research, we show that China has topped in the recent two decades. Singapore and Hong Kong also exhibit rapid increases in their representation in atypical science.

\subsection{Contribution to the literature}

Our work speaks to a number of related literature. First, our finding that propensity to produce novel research continues to vary by country is in line with existing studies on cross-nation comparisons of productivity and scientific impact as measured by citations \citep{narin1975national, may1997scientific, horta2007opening, King-nation-2004, XieKillewald2012}. However, some of our findings differ from these previous studies. Particularly, analyses on the tendency to publish highly cited papers reveal the US as a long-time leader that is at the head of other large producers like UK and Germany and remains even further ahead of China and Singapore. By contrast, our analysis on the tendency to introduce novel combinations of knowledge elements indicates that China and Singapore have become leaders that outpace the US and many western European countries. Moreover, some prolific countries like UK and Canada lagged far behind in terms of scientific novelty.

The case of China needs some attention. Many studies have pointed out the rapid rise of China as a major scientific and technological powerhouse \citep{xie2014china, ZHOU200683, xie2019bigger}. They found that not only China has become one of the largest producers in science but also its share in top highly-cited papers has been increasing to an extent that may be in an equally important position as the US. Our analysis adds to this line of literature by revealing that China has also emerged as a country that disproportionately produced more novel and atypical research than the US did.

The differentiated ranking in productivity, impact, and novelty is not unexpected. After all, extant studies have shown that novelty is correlated with impact, but only partially \citep{WANG20171416}. Novelty is an ex ante characteristic that can be early independent proxy to identify scientific breakthrough and recognize their importance in time, whereas impact is ex post measure that is dependent on numerous factors like author reputation, institution and venue prestige.

From the input-output perspective, our results are consistent with previous inquiry that found large differences in nations' capabilities in goods production as measured by exports \citep{hidalgo2007product, hidalgo2009building}. A commonality in these analyses is that capabilities affect the types of inputs a country uses.

A few lines of existing literature may help theorize specific mechanisms driving differences in the tendency to publish novel research, though beyond the scope of this work. The first is whether international collaboration may lead to scientific novelty. However, \citet{Wagner-intl-2019} explored the relationship between novelty and international collaboration, finding that international collaboration is associated with conventional rather than novel or atypical research. This association implies that domestic collaboration tends to produce novel research, raising the question of which countries are more likely to have more novel science published, which is the focus of our work.

Although international collaboration may incur ``transaction costs and communication barriers'', which may suppress novel science production, collaboration still remains a viable way for novel knowledge production. This is because novel science explores uncharted territories, which may be more dependent on interactions with colleagues with complementary knowledge so that a synergistic effect can be realized. In this regard, cross-country variation in the tendency to produce novel ideas may stems from fruitful domestic interactions.

\subsection{Policy implications}

The ability to evaluate a nation's standings in science and technology is of great concern to its government and other relevant stakeholders, as the outcome of such evaluations has a direct influence on policy and funding. Ever since the birth of the citation index database, the focus in science and technology indicators for knowledge production has been on productivity and (citation-based) impact, which are greatly emphasized in funding agencies' reports like the \emph{Science and Engineering Indicators} from the NSF. As such, science policy-making has been shaped, at least partially, by considering the impact of scientific work. A potential unintended consequence of the excessive stress on impact is that researchers have become more risk-averse and science has become less vibrant, a phenomenon that has already been observed in biomedicine \citep{Rzhetsky14569}. Many have alarmed that the scientific community's obsession with impact metrics may have led to stagnant science \citep{alberts2013impact, Verma2015Impact}.

The novelty indicators introduced here may possess potential for science policy, by providing a more comprehensive array of indicators beyond impact that can serve as complementary tools in science evaluations and policy making. As novel ideas are raw, their maturity into meaningful scientific advances needs attention and further experimentation by many scientists. However, this kind of research is risky, and in the absence of explicit incentives, many scientists may prefer to pursue more conventional research that is already well explored by others, with an expectation for receiving more citations---fitting into the existing framework in science policy. By using novelty measures in conjunction with traditional impact-based indicators, funding agencies and university administrators can re-structure their reward systems, striking a better balance between rewarding research that exploits mature, well-established ideas and rewarding research that explores and develops ideas at early stages. While a singular emphasis on impact under-reward scientists who try out new ideas, rewarding scientists who pursue more novel research paths would encourage them to strive for scientific novelty, thereby leading to the formation of a scientific community that is vibrant to new venues for scientific inquiry.

To be sure, novelty metrics are not meant to replace traditional impact ones. Instead, as impact and novelty capture distinct aspects of science, complementarity is our suggestion here---scientists should be rewarded based on the impact and novelty of their work. A related note is that novel work is known to be more likely to be highly influential in the traditional sense of receiving a large number of citations, but oftentimes not within a short time-window. Moreover, novel research can have influence that is not readily captured by citations; it generates scientific value by trying out new possibilities of idea combinations and paving the ways for their maturity.

\subsection{Limitations and future work}

Our work has the following limitations. First, a significant portion of papers were excluded from our analysis of cross-nation comparison of the production of novel and atypical science, due to the lack of affiliation data. Second, we adopted pairwise combinatorial novelty, viewing novel research as results from combining knowledge components in an unprecedented fashion and treating journals as knowledge components. There may be alternative entities like keywords and topics to be identified as knowledge elements and other possible lines of thought to define novelty, such as novelty from network structure perspective. Third, our focus in this work was on nations, and it would be useful to extend the analysis to cities to examine the concentration of novel science in large cities and compare the extent with other innovative activities. This would contribute to the literature on agglomeration economies.

Fourth, we have yet to reveal any specific mechanisms driving the propensity to pursuit novel science. Future work towards this goal could be from several aspects. For example, an important direction is to study determinants of novel idea adoption of researchers. Do team composition, mentorship, and researchers' demographics and career stage play a role? At the science funding level, one could also investigate strategies for effective funding allocations. Is a diversified schema for resource allocation more effective in terms of collective scientific discovery? Finally, a joint modeling of novelty in science and technology would deepen our understanding of the relationships between scientific and technological frontiers, revealing, for example, to what extent scientists with novel publications are also novel inventors, whether technological breakthroughs build on both novel science and novel technology, and so on. In all these regard, our release of novelty and conventionality classifications of papers in a large-scale bibliographic database may open up venues for further systematic studies of creativity in science.

\section*{Acknowledgments}

We thank Alexander J. Gates for useful discussions. QK is partially supported by the National Natural Science Foundation of China (72204206), City University of Hong Kong (Project No. 9610552, 7005968), and Hong Kong Institute for Data Science.

\clearpage
\includepdf[pages=-]{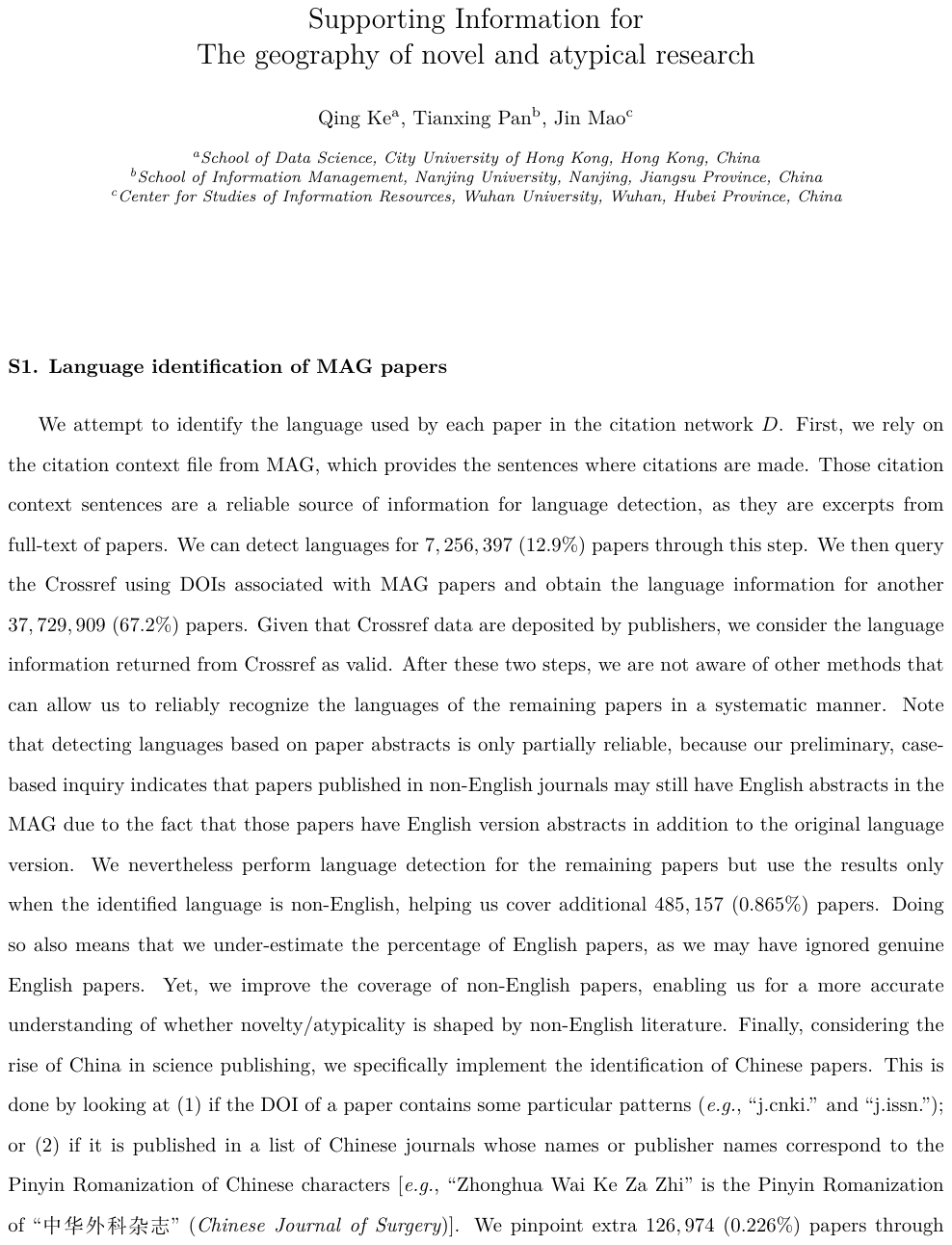}

\end{document}